\documentclass[prb,showpacs,twocolumn,amsmath,amssymb,superscriptaddress]{revtex4}
\usepackage{graphicx}
\usepackage[colorlinks,bookmarks=false,citecolor=red,linkcolor=blue,urlcolor=blue]{hyperref}
\newcommand{\mbk}{\mathbf{k}}

\def\r{{\bf r}}     
\def\n{{\bf n}}    
\def\Sbf{\textbf{S}}
\def\Qbf{\textbf{Q}}
\allowdisplaybreaks

\begin{document}

\title{Extended degeneracy and order by disorder in the square lattice J$_1$-J$_2$-J$_3$ model}
\author{Bimla Danu}
\email{bimladanu@imsc.res.in}
\affiliation{The Institute of Mathematical Sciences, C I T Campus, Chennai 600 113, India}
\author{Gautam Nambiar}
\affiliation{The Institute of Mathematical Sciences, C I T Campus, Chennai 600 113, India}
\affiliation{Indian Institute of Science, Bangalore 560012, India}
\author{R. Ganesh}
\email{ganesh@imsc.res.in}
\affiliation{The Institute of Mathematical Sciences, C I T Campus, Chennai 600 113, India}

\date{\today}

\begin{abstract}
The square lattice antiferromagnet with frustrating next nearest neighbour coupling continues to generate tremendous interest, with an elusive quantum disordered phase in the vicinity of $J_2$ = $J_1$/2. At this precise value of frustration, the classical model has a very large degeneracy which makes the problem difficult to handle. We show that introducing a ferromagnetic $J_3$ coupling partially lifts this degeneracy. It gives rise to a four-site magnetic unit cell with the constraint that the spins on every square must add to zero. 
 This leads to a two-parameter family of ground states and an emergent vector order parameter. We reinterpret this family of ground states as coexistence states of three spirals.
Using spin wave analysis, we show that thermal and quantum fluctuations break this degeneracy differently. Thermal fluctuations break it down to a threefold degeneracy with a N\'eel phase and two stripe phases. This threefold symmetry is restored via a $Z_3$ thermal transition, as we demonstrate using classical Monte Carlo simulations. On the other hand, quantum fluctuations select the N\'eel state. In the extreme quantum limit of spin-$1/2$, we use exact diagonalization to demonstrate N\'eel ordering beyond a critical $J_3$ coupling. 
For weak $J_3$, a variational approach suggests an $s$-wave plaquette-RVB state. 
Away from the $J_2 = J_1/2$ line, we show that quantum fluctuations favour N\'eel ordering strongly enough to stabilize it within the classical stripe region. Our results shed light on the origin of the quantum disordered phase in the $J_1$-$J_2$ model.
\end{abstract}
                                 
\maketitle

\section{Introduction}
The paradigmatic example of frusrated magnetism is the square lattice antiferromagnet with next-nearest neighbour coupling: the $J_1 $-$J_2$ model. It is well known that it has N\'eel antiferromagnetic order when $J_2 \ll J_1$ and stripe order when $J_2\gg J_1$. The effects of frustration become apparent in the intermediate regime when $J_2 \sim  J_1/2$. The nature of the quantum ground state in this regime continues to be debated with several proposals for plaquette order~\cite{Capriotti2000,Gong2014,Yu2012}, a valence bond crystal~\cite{Dhuse1988,Gelfand1989,Sachdev1990,NRead1991,Poilblanc1991,Singh1999,Kotov1999,Morita2015}, etc. Notably, there are several proposals for a spin liquid with topological order~\cite{Balents2012,Capriotti2001,Morita2015}. 

The complex and rich behaviour that intervenes between the N\'eel and stripe ground states has its origin in the classical spin model. Precisely at $J_2 = J_1/2$, the classical phase boundary between N\'eel and stripe ground states, the classical problem has an extensively degenerate ground state manifold~\cite{Chandra1990}. 
Quantum fluctuations can select correlations from within this manifold to form various ordered phases. Indeed, this is the underlying reason behind the many competing claims about the quantum $S=1/2$ phase diagram. While this degeneracy gives rise to a rich phase diagram, it makes it extremely difficult to understand this parameter regime. In this paper, we make the problem tractable by introducing a suitable tuning knob -- a ferromagnetic third-neighbour coupling. This $J_3$ coupling partially lifts the degeneracy of the $J_2 = J_1/2$ problem; it does so in an elegant and tunable manner that allows for an understanding of 
the classical and quantum phase diagrams.

The extended degeneracy in the problem at hand occurs at a classical phase boundary. It is well known that extended ground state degeneracies may occur at phase boundaries~\cite{Rosales2013, Yamamoto2015, Seabra2016}. Here, the residual degeneracy after introducing $J_3$ is given by a local constraint that leads to a four-site magnetic unit cell. Equivalently, it can be understood in terms of coexisting spiral states. Similar physics has recently been seen in the honeycomb lattice $J_1$-$J_2$ problem, where a magnetic field is used to select different combinations of spirals~\cite{Rosales2013}. 

The rest of this paper is organized as follows. Section~\ref{sec.classical} describes the classical phase diagram of the $J_1$-$J_2$-$J_3$ problem, bringing out the special role of a ferromagnetic $J_3$ interaction. Section~\ref{sec.coexisting} shows why coexisting spirals are allowed ground states for the parameters of interest, and how they give rise to an extensive degeneracy. Sections~\ref{sec.sumsquares},~\ref{sec.sumsquaresJ3} present the ground state degeneracy as a local constraint on every square plaquette. Sections~\ref{sec.largeS} and~\ref{sec.thermal} describe the breaking of the classical degeneracy by weak quantum and thermal fluctuations respectively. Section~\ref{sec:classicalMC} describes classical Monte Carlo results that establish a thermal $Z_3$ transition. Section~\ref{sec.ED} addresses the $S=1/2$ limit, with~\ref{subsec:ED} discussing exact diagonalization results,~\ref{subsec:stabNeel} discussing the stabilization of N\'eel order into the stripe domain and~\ref{subsec:variational} presenting a variational plaquette wavefunction. Finally, section~\ref{discussion} summarizes our results and discusses consequences for the quantum disordered phase in the $J_1$-$J_2$ problem. 

\section{Classical phase diagram}
\label{sec.classical}
The Heisenberg model on the square lattice is well known as the parent Hamiltonian of the undoped cuprates\cite{Efstratios1991}. We study an extended version of this Hamiltonian given by
\begin{equation}
H=J_{1}\sum_{\langle i,j\rangle}\Sbf_i.\Sbf_j+J_{2}\sum_{\langle\langle i,j\rangle\rangle}\Sbf_i.\Sbf_j+J_{3}\sum_{\langle\langle\langle i,j\rangle\rangle\rangle}\Sbf_i.\Sbf_j,
\label{energy_expression}
\end{equation}
where  $\langle i,j \rangle$, $\langle\langle i,j \rangle\rangle$ and $\langle\langle\langle i,j \rangle\rangle\rangle$ refer to nearest neighbours, next-nearest neighbours, and third nearest neighbours, respectively. We take the couplings $J_1$ and $J_2$ to be antiferromagnetic. Choosing $J_3$ to be ferromagnetic leads to interesting consequences as we argue below.

\subsection{Method of spiral states}
\begin{figure}
\includegraphics[width=85mm]{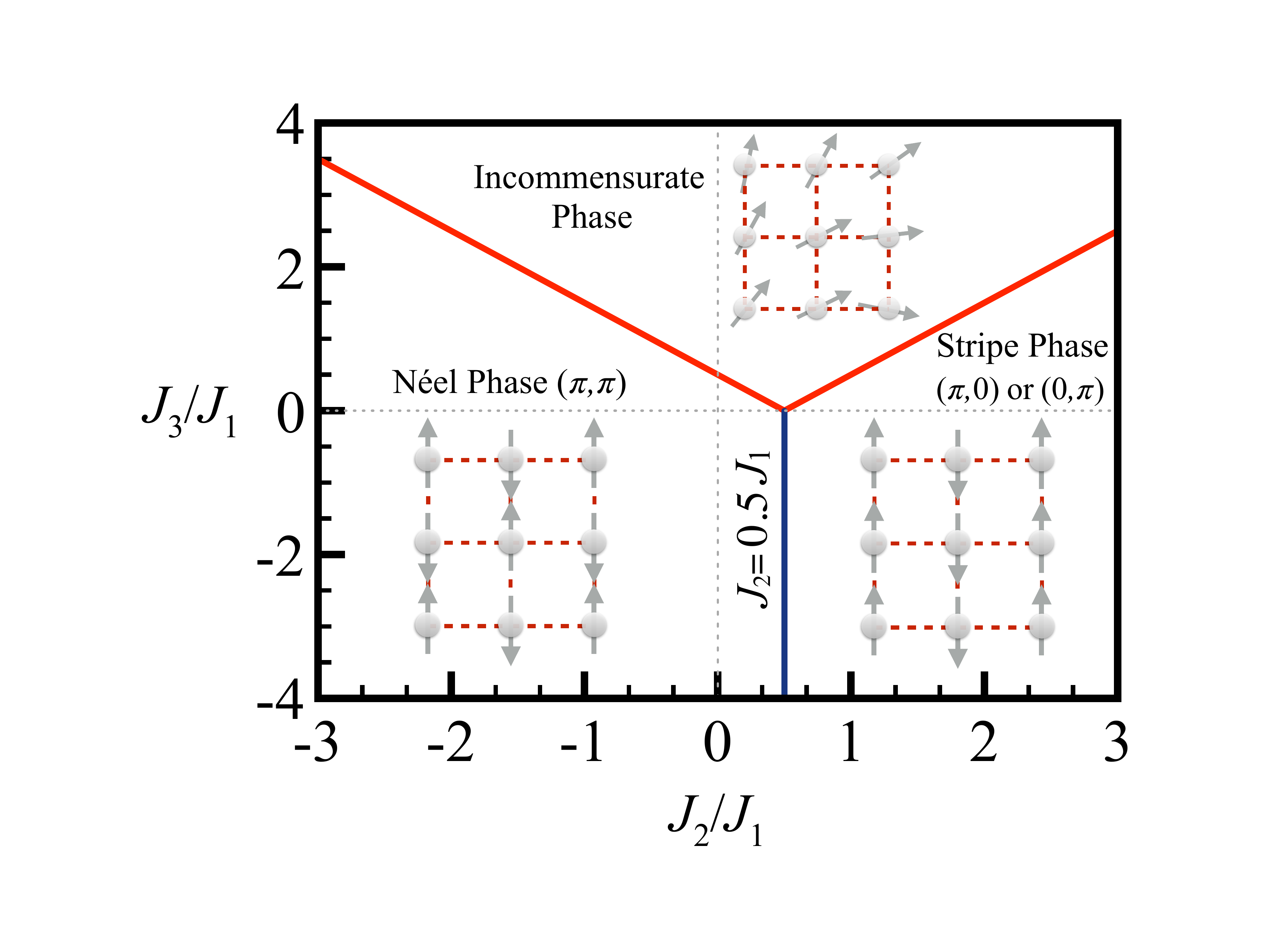}
\caption{Classical phase diagram with antiferromagnetic $J_{1}$. N\'eel and stripe phases are separated by the line $J_2 = J_1/2$, $J_3\leq 0$. The incommensurate phase is bounded by the lines $J_3=0.5 \lvert J_2-0.5 J_1\rvert$.}
\label{J2J3graph}
\end{figure}

To find the classical ground state for given $J_1$, $J_2$ and $J_3$, we use the method of spiral states\cite{Villain1977,Fouet2001,Mulder2010}.  As a variational ansatz, we define a coplanar spiral characterized by a pitch vector $\mathbf{Q}$,
\begin{equation}
\textbf{S}_i=S\{ \cos{(\textbf{Q}.\textbf{r}_i)}\hat{x} + \sin{(\textbf{Q}.\textbf{r}_i)}\hat{y} \}.
\label{eq.spiral}
\end{equation}
This state breaks spin rotational symmetry spontaneously. We have chosen the XY plane for concreteness; the ordering could occur in any plane.   
The energy of this state is given by 
\begin{eqnarray}
\nonumber E_{\mathbf{Q}}/NS^2 &=& J_1 (\cos{Q_x}+\cos{Q_y}) + 2J_2 \cos{Q_x}\cos{Q_y}\\
 &+& J_3 (\cos{2Q_x}+\cos{2Q_y}),
\end{eqnarray}
 where $N$ is the total number of spins. Minimizing with respect to $\mathbf{Q}$, we obtain the classical phase diagram shown in Fig.~\ref{J2J3graph}. There are three well-defined regions: N\'eel, stripe and incommensurate. In the N\'eel region, the ground state is the standard N\'eel antiferromagnet with $\mathbf{Q}=(\pi,\pi)$. The stripe phase breaks a $Z_2$ symmetry corresponding to the choice between horizontal and vertical stripe order\cite{Chandra1990}. The ordering wavevector is $\mathbf{Q} = (0,\pi)$ or $(\pi,0)$. In both N\'eel and stripe phases, the wavevector $\mathbf{Q}$ is fixed at high-symmetry points on the Brillouin zone edge. In contrast, in the incommensurate phase, the value of $\mathbf{Q}$ changes with the coupling strengths\cite{Mambrini2006}. The incommensurate phase has been shown to give rise to a quantum non-magnetic phase along one particular line in the space of couplings\cite{Mambrini2006}. While this phase diagram has been extensively studied for antiferromagnetic $J_3$~\cite{Gelfand1989,Richter1997,Hauke2011,Reuther2011,Mikheenkov2011,Mezio2013}, we focus on the case of ferromagnetic $J_3$ here. 
 A similar phase diagram has been found for ferromagnetic $J_1$~\cite{Sindzingre2009}.

\section{Extended degeneracy along the ($J_2=J_1/2,J_3 < 0$)  line}

The line defined by $J_3 < 0$ and  $J_2 = J_1/2$ is the phase boundary between N\'eel and stripe phases. Na\"ively, we may expect that the classical ground state here to be three fold degenerate with N\'eel, horizontal stripe and vertical stripe ground states. However, the degeneracy is much larger as we show below. 

\begin{figure}
\includegraphics[width=65mm]{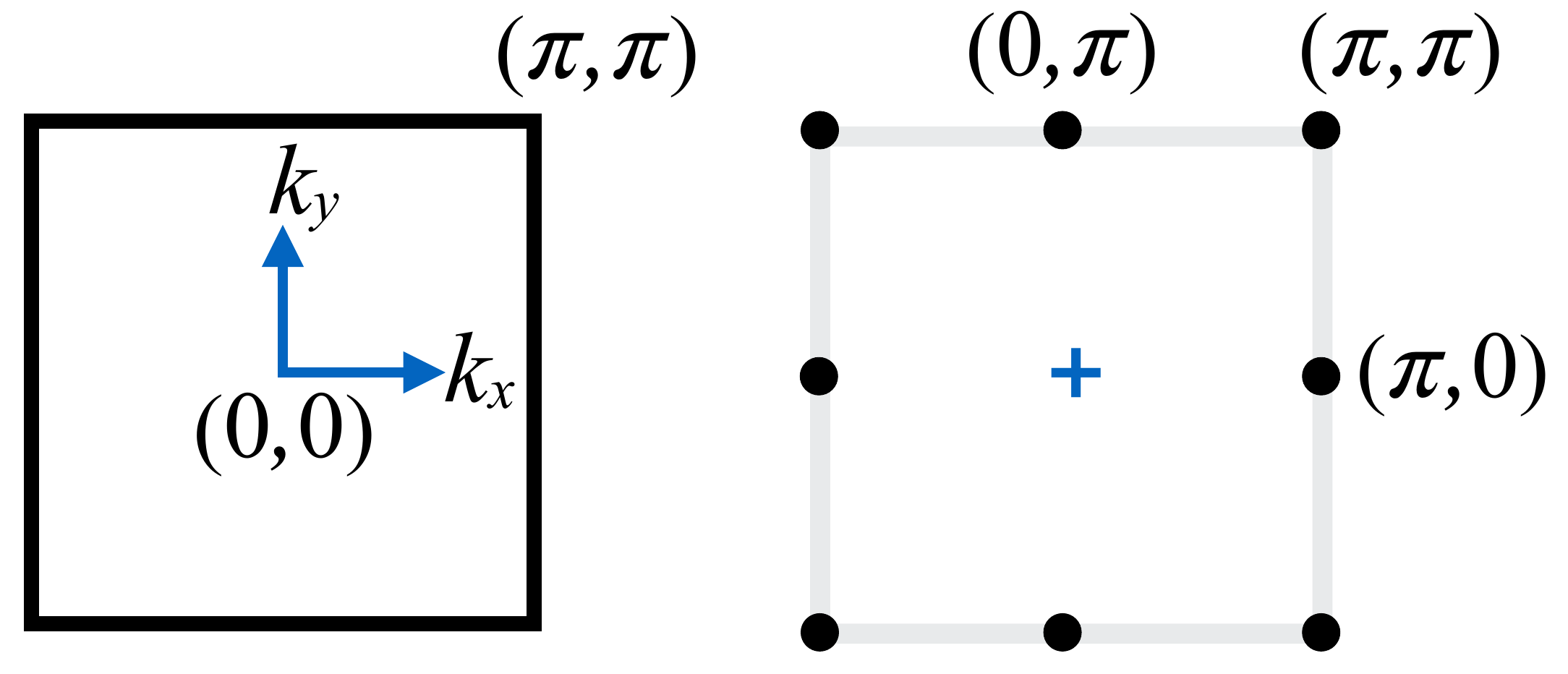}
\caption{Left: Ground state spiral wavevectors for $(J_2 = J_1/2; J_3 = 0)$. Right: For  $(J_2 = J_1/2; J_3 < 0)$.}
\label{fig:classminima}
\end{figure}

\subsection{Coexisting Spirals}
\label{sec.coexisting}

At $(J_2 = J_1/2, J_3 =0 )$, the method of spirals gives an infinitely degenerate ground state. Minimizing the variational energy picks all $\mathbf{Q}$'s that lie on the edge of the Brillouin zone, as shown in Fig.~\ref{fig:classminima}(left). A ferromagnetic $J_3$ breaks this degeneracy and picks three wavevectors as shown in Fig.~\ref{fig:classminima}(right): $\mathbf{Q}_1=(\pi,\pi)$ corresponding to N\'eel, $\mathbf{Q}_2=(0,\pi)$ corresponding to horizontal stripe and $\mathbf{Q}_3=(\pi,0)$ corresponding to vertical stripe ordering. All three $\mathbf{Q}$'s satisfy the special property of being half a reciprocal lattice vector, i.e., $2\mathbf{Q} \equiv 0$. As shown by Villain~\cite{Villain1977}, this property allows the spirals to coexist. To show this, we first note that the three $\mathbf{Q}$'s satisfy $\sin({\mathbf{Q}\cdot \mathbf{r}_i}) = 0$ at every lattice point. Therefore, in a spiral state as in Eq.~\ref{eq.spiral}, we may only retain the cosine terms. 
A coexisting spiral can be written as 
\begin{equation}
\textbf{S}_i=S \{ \cos{(\textbf{Q}_1.\textbf{r}_i)}\hat{u} + \cos{(\textbf{Q}_2.\textbf{r}_i)}\hat{v}  + \cos{(\textbf{Q}_3.\textbf{r}_i)}\hat{w} \},
\label{eq.coexistingspiral}
\end{equation}
where $\hat{u}$, $\hat{v}$, $\hat{w}$ are arbitrary vectors. This is an allowed spin configuration if the spin length is preserved at every site. This condition gives us the following constraints, upon using the properties of $\Qbf_{1,2,3}$:
\begin{eqnarray}
 \nonumber \vert \hat{u} \vert^2 + \vert \hat{v} \vert^2 + \vert \hat{w} \vert^2 = 1, \\
 \hat{u} \cdot \hat{v} =  \hat{v} \cdot \hat{w} = \hat{w} \cdot \hat{u} = 0.
 \label{eq.coexistconditions}
 \end{eqnarray}
We note that the ability to form coexisting spirals is a special feature of the ($J_2 = J_1/2$, $J_3 < 0$) line. For example, the incommensurate phase in Fig.~\ref{J2J3graph} does have multiple $\mathbf{Q}$ solutions. However, they cannot be combined into a coexisting state with uniform spin length. 

The state in Eq.~\ref{eq.coexistingspiral} has nine independent parameters -- three components each of $\hat{u}$, $\hat{v}$ and $\hat{w}$. After taking into account the four constraints in Eqs.~\ref{eq.coexistconditions}, we have five degrees of freedom in choosing the ground state. From the three $\mathbf{Q}$'s, it is easy to see that the coexistence state in Eq.~\ref{eq.coexistingspiral} has a four-site unit cell. The allowed ground states and the unit cell can also be understood from a local constraint as we show below.  

\subsection{Sum of squares argument with $J_3=0$}
\label{sec.sumsquares}
\begin{figure}[htbp]
\centering
\includegraphics[width=85mm]{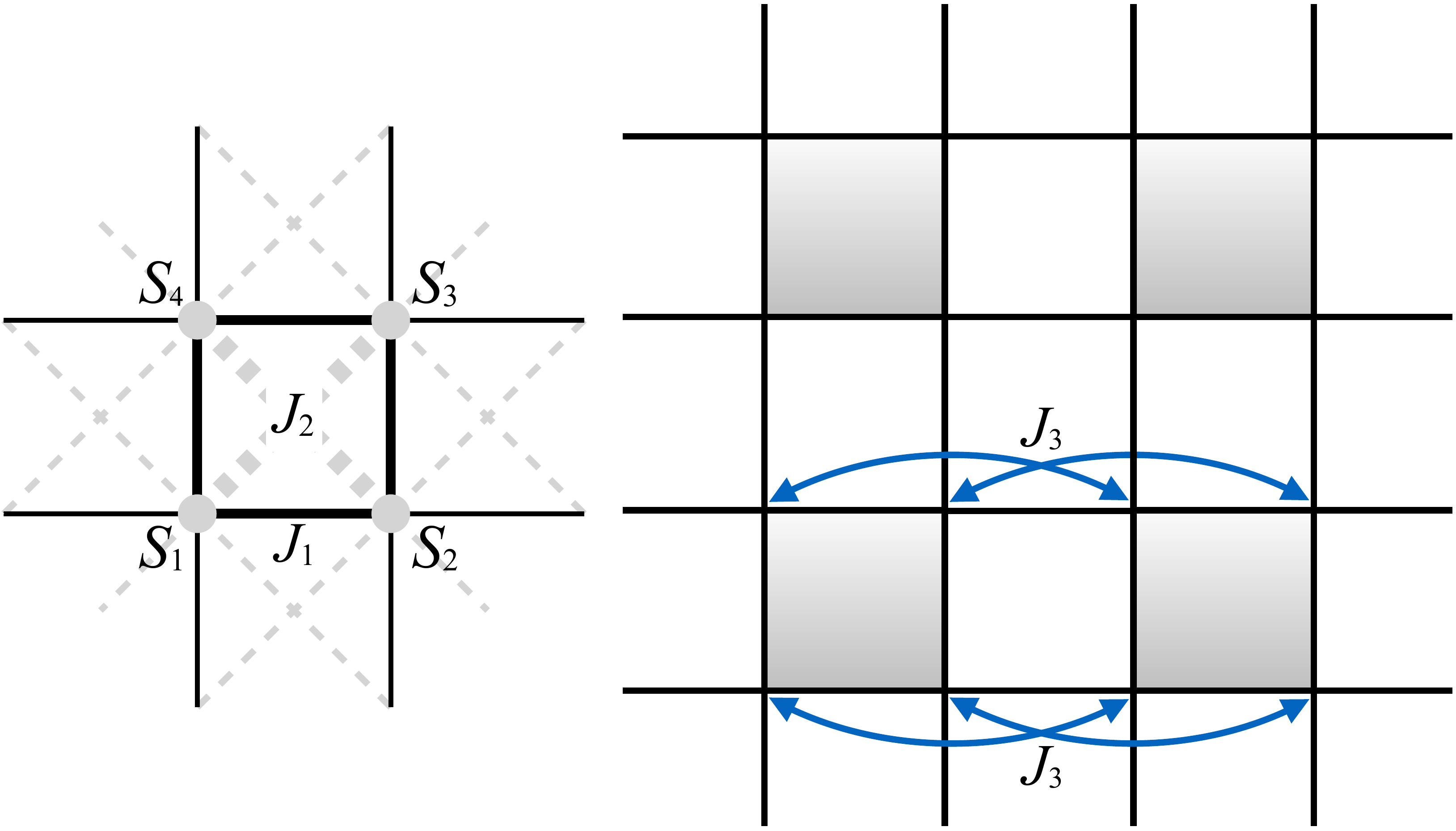}
\caption{Left: When $J_3=0$, the Hamiltonian can be written as a sum over squares. Each $J_1$ bond is shared between two adjacent squares, while each $J_2$ bonds only occurs in one square. Right: Shaded squares represent the magnetic unit cell favoured by $J_3$: the ferromagnetic $J_3$ bonds ensure that all shaded squares have the same spin configuration.
}
\label{fig.squares}
\end{figure}

Let us first consider the $J_3 = 0$ case with $J_2 = J_1/2$. At this special point, the classical Hamiltonian can be written as a sum over squares~\cite{Chandra1990},
\begin{equation}
H_{J_3 = 0} = \sum_{\boxtimes}H_\boxtimes = \sum_{\boxtimes} \frac{J_1}{4} (\mathbf{S}_1 + \mathbf{S}_2 + \mathbf{S}_3 + \mathbf{S}_4)^2,
\end{equation}
where the sum is over every square plaquette -- see Fig.~\ref{fig.squares}(left). 
The decomposition into a sum over squares works because each $J_1$ bond is shared between two adjacent squares, while each $J_2$ bond only appears in one square. As the Hamiltonian is a sum over positive quantities, the ground state is given by the condition that each square should have zero total spin, i.e., 
\begin{equation}
\mathbf{S}_1 + \mathbf{S}_2 + \mathbf{S}_3 + \mathbf{S}_4 =0,
\label{eq.zerosum}
\end{equation}
on every square. As we argue below, this local constraint leads to an infinitely degenerate ground state manifold reminiscent of spin-ice~\cite{Gingras2011}. 

\begin{figure}[htbp]
\centering
\includegraphics[width=8.6cm]{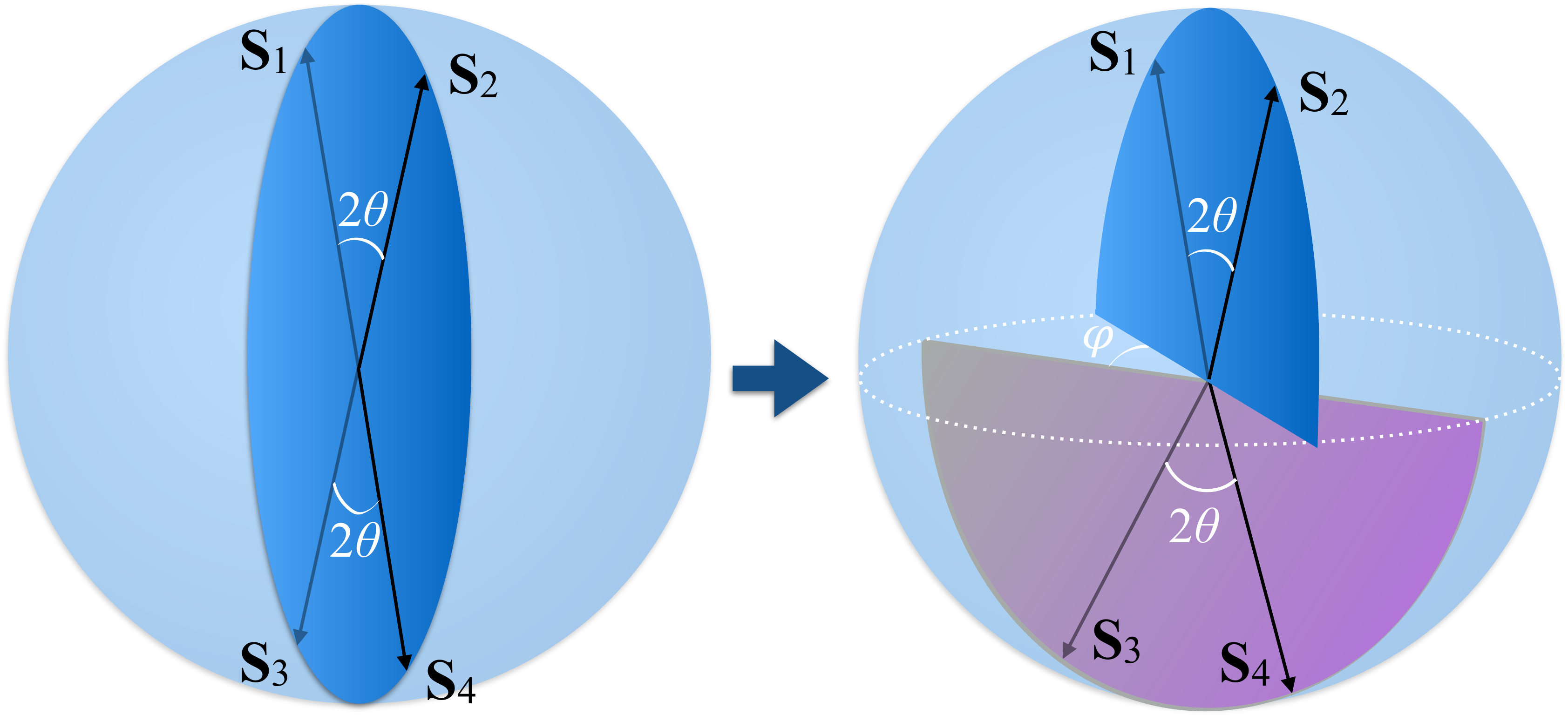}
\caption{Parametrizing the ground of a single square with a zero-total-sum constraint by two angles. We first take all spins to lie in one plane so that $\mathbf{S}_1$ and $\mathbf{S}_2$  make an angle $2\theta$. 
We choose $\mathbf{S}_3 = -\mathbf{S}_2$ and $\mathbf{S}_4 = -\mathbf{S}_1$ to satisfy the zero-total-spin constraint. We then rotate $\mathbf{S}_3$ and $\mathbf{S}_4$ about the $\mathbf{S}_1+\mathbf{S}_2$ axis by an angle $\varphi$.
}
\label{fig.thetaphi}
\end{figure}

Let us first consider a single square. An allowed spin configuration is given by a choice of four vectors on the Bloch sphere which satisfy Eq.~\ref{eq.zerosum}.
Such a configuration can be described by two angles $\theta$ and $\varphi$, upto an overall spin rotation. As depicted in Fig.~\ref{fig.thetaphi}, $\mathbf{S}_1$ and $\mathbf{S}_2$ are initially chosen to make an angle $2\theta$ with each other. The spins $\mathbf{S}_3$ and $\mathbf{S}_4$ are chosen to lie on the same plane with $\mathbf{S}_3=-\mathbf{S}_2$ and  $\mathbf{S}_4=-\mathbf{S}_1$, thereby satisfying the zero-total-spin condition. We have one more degree of freedom in rotating $\mathbf{S}_3$ and $\mathbf{S}_4$ about the $\mathbf{S}_1 + \mathbf{S}_2$ axis by the angle $\varphi$. With this parametrization, taking $\hat{z}$ to be parallel to $\mathbf{S}_1 + \mathbf{S}_2$, we arrive at
\begin{eqnarray}
\left[\mathbf{S}_1, \mathbf{S}_2, \mathbf{S}_3, \mathbf{S}_4 \right] \!= \! S \!
\left[ \!\hat{n}_{\{\theta,0\}}, \hat{n}_{\{\theta,\pi\}}, \hat{n}_{\{\pi-\theta,\varphi\}}, \hat{n}_{\{\pi-\theta,\varphi+\pi\}}\!\right]\!\!,
\end{eqnarray}
where $\hat{n}_{\{\alpha,\beta\}}$ denotes a unit vector with polar angle $\alpha$ and azimuthal angle $\beta$. We assert that any spin configuration on a square that satisfies Eq.~\ref{eq.zerosum} can be obtained by a suitable choice of $\{\theta,\varphi\}$ followed by a global spin rotation.

On the full two-dimensional square lattice, the problem of enumerating all allowed ground states reduces to that of assigning $\{\theta,\varphi\}$ to each square, keeping in mind that neighbouring squares are coupled. It is easy to see that this leads to an infinite number of ground state configurations. We note here that the domain of $\theta$ is $[0,\pi]$, while that of $\varphi$ is $[0,2\pi)$; the parameters $\{\theta,\varphi\}$ thus define an emergent vector field with unit length. An effective field theory for the $J_3=0$ problem would involve a vector field with fixed length coupled to an $SO(3)$ matrix field that encodes spin rotations.

\subsection{Sum of squares argument with $J_3<0$}
\label{sec.sumsquaresJ3}
Introducing a ferromagnetic $J_3$ coupling leads to a drastic simplification. As shown in Fig.~\ref{fig.squares}(right), the $J_3$ term forces every alternating square to have the same spin configuration. The ground state is completely fixed once we fix $\mathbf{S}_1$, $\mathbf{S}_2$, $\mathbf{S}_3$ and $\mathbf{S}_4$ on one shaded square. Moreover, if the spins on the shaded square are chosen to satisfy Eq.~\ref{eq.zerosum}, the unshaded squares automatically satisfy Eq.~\ref{eq.zerosum} as well. Such a spin configuration will minimize the $J_1$-$J_2$ energy contribution, while maximally lowering its energy from the $J_3$ bonds. 

Thus, with a ferromagnetic $J_3$ coupling, all possible ground states are obtained by constraining $\mathbf{S}_i$'s on one square so as to satisfy Eq.~\ref{eq.zerosum}. This gives us a two-parameter ground state manifold (upto global spin rotations) characterized by $\{\theta,\varphi\}$ or equivalently by a vector of unit length. With three Euler angles required to define a global spin rotation matrix, we have five degrees of freedom in total -- in agreement with the coexisting spirals argument in Section \ref{sec.coexisting}.

\section{Spin wave analysis}

We have established that the classical model with $J_2 = J_1/2$ and $J_3<0$ has a two parameter ground state manifold. This degeneracy can be broken by thermal/quantum fluctuations by the well-known `order by disorder' mechanism~\cite{villain1980}. To demonstrate this, we consider spin wave fluctuations about a generic state in the ground state manifold.

\begin{figure}[htbp]
\centering
\includegraphics[width=0.5\textwidth]{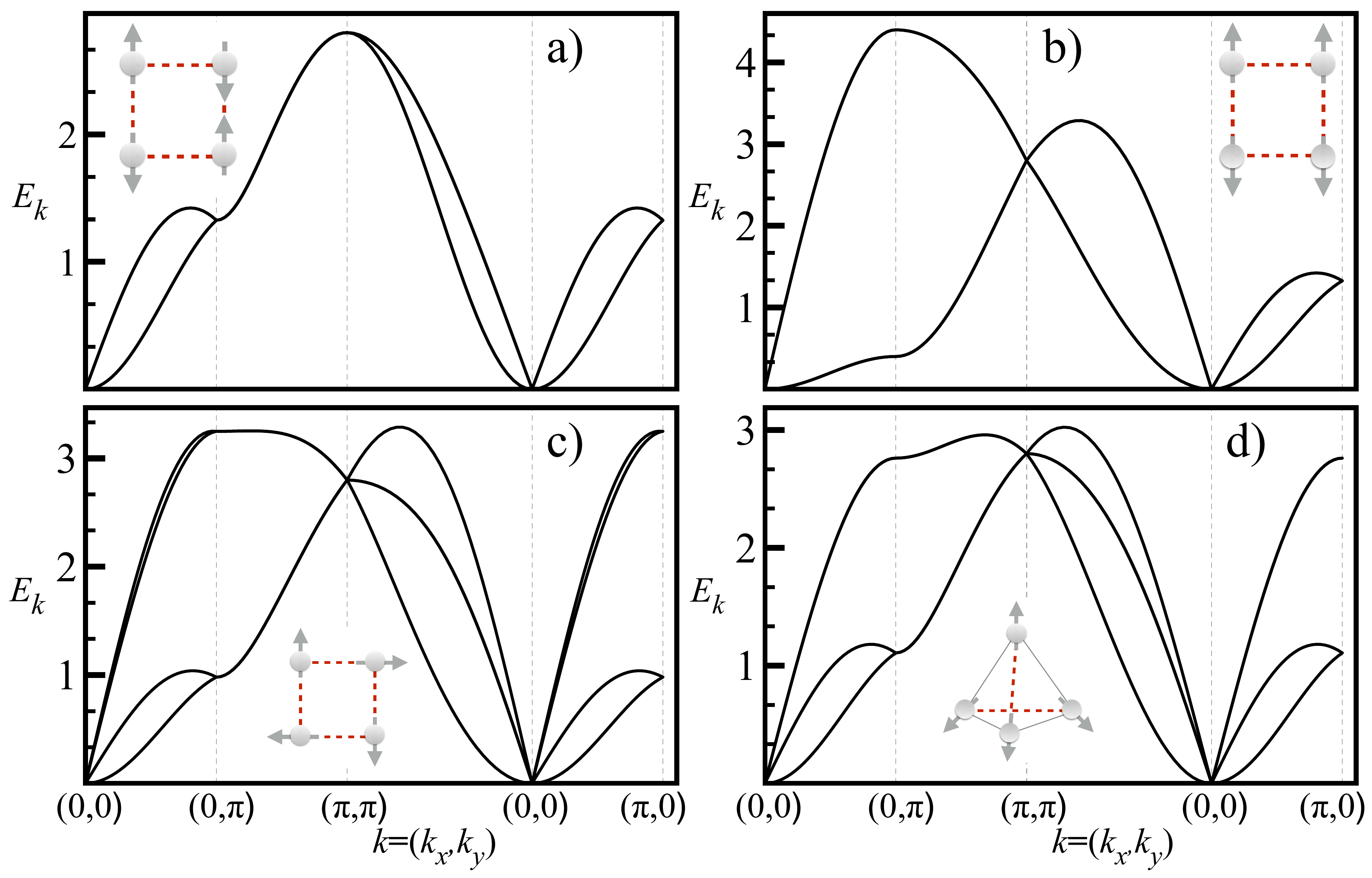}
\caption{Spin wave dispersion of four possible ground states: a) N\'eel, b) Stripe, c) Coplanar and d) tetrahedral (non-coplanar) states. The schematic in each panel shows the four spins in the magnetic unit cell. In all the ground states, there are goldstone modes which go to zero linearly as well as those that go to zero quadratically.}
\label{fig:spinwavedisp}
\end{figure}

As argued above, all the allowed ground states have a four-site magnetic unit cell. Performing the usual Holstein Primakov transformation and retaining $\mathcal{O}(S)$ terms, we obtain a quadratic Hamiltonian of the form
\begin{equation}
H_{\mathcal{O}(S)} = -8J_3N_{\boxtimes}S^2+ \sum_{\mathbf{k}} {}^{'}
\left(  \begin{array}{cc}
\psi_\mbk^\dagger & \psi_{-\mbk} 
\end{array} \right)
H_{8\times 8} (\mbk)
\left(  \begin{array}{c}
\psi_\mbk \\
 \psi_{-\mbk}^\dagger.
\end{array} \right).
\end{equation}
The sum is over half the Brillouin zone and $N_{\boxtimes}$ is the number of unit cells in the system -- shaded squares in Fig.~\ref{fig.squares}(right). We have denoted $\psi_\mbk^\dagger
= \{\begin{array}{cccc} a_{1,\mbk}^\dagger & a_{2,\mbk}^\dagger & a_{3,\mbk}^\dagger& a_{4,\mbk}^\dagger \end{array}\}$, where $a_{1,\mbk}^\dagger$ creates a spin wave fluctuation with momentum $\mbk$ on the sublattice $i$. The $8\times8$ matrix with $\mathcal{O}(S)$ terms can be diagonalized by a bosonic Bogoliubov transformation to give
\begin{eqnarray}
\nonumber H_{\mathcal{O}(S)} &=&  -8J_3N_{\boxtimes}S^2 +\\
&{}& \sum_{\mathbf{k}}{}^{'} \!\sum_{j=1}^4 \!\epsilon_{j,\mbk} \{ \gamma_{j,\mbk}^\dagger \gamma_{j,\mbk} \!+ \! \gamma_{j,-\mbk}\gamma_{j,-\mbk}^\dagger  \} \! + \! c_\mbk,
\end{eqnarray}
where $ \epsilon_{j,\mbk}$ are the spin wave energies,  $c_\mbk$ is a $\mbk$-dependent constant and $ \gamma_{j,\mbk}^\dagger$ is the eigenmode  creation operator. 
In Fig.~\ref{fig:spinwavedisp}, we illustrate the spin wave spectrum for four possible ground states. We have chosen four highly symmetric configurations for the purpose of illustration: N\'eel, stripe, coplanar and tetrahedral orders. 

As in the four states in Fig.~\ref{fig:spinwavedisp}, we find two kinds of Goldstone modes in all allowed ground states: linear modes with $\epsilon_{j,\mbk}\sim k$ as well as quadratic modes with $\epsilon_{j,\mbk}\sim k^2$. Linear modes usually occur in antiferromagnets while quadratic modes occur in ferromagnets. Our system combines both these elements. 

\subsection{Quantum order by disorder}\label{sec.largeS}
\begin{figure}[htbp]
\centering
\includegraphics[width=4.25cm]{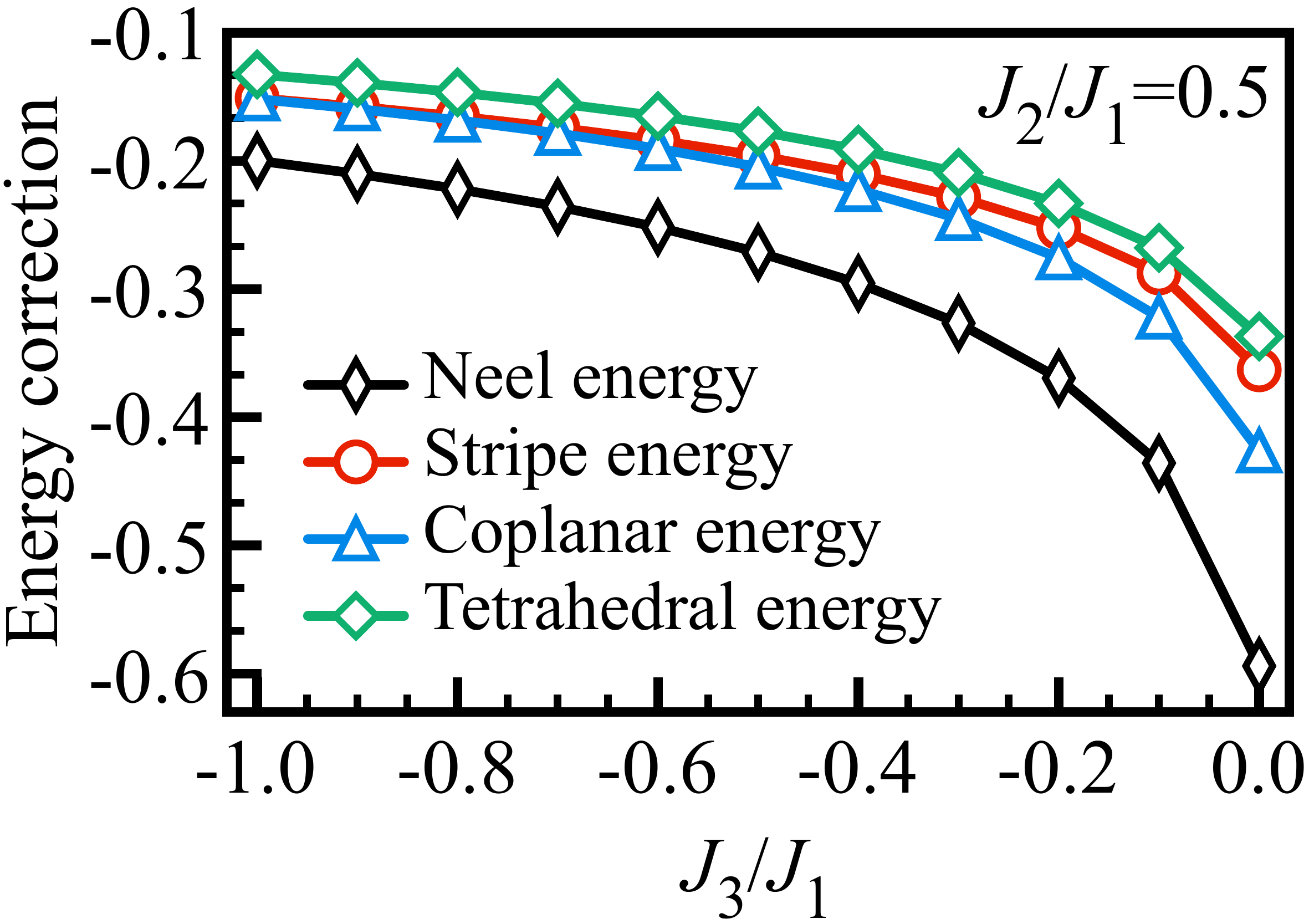}
\includegraphics[width=4.25cm] {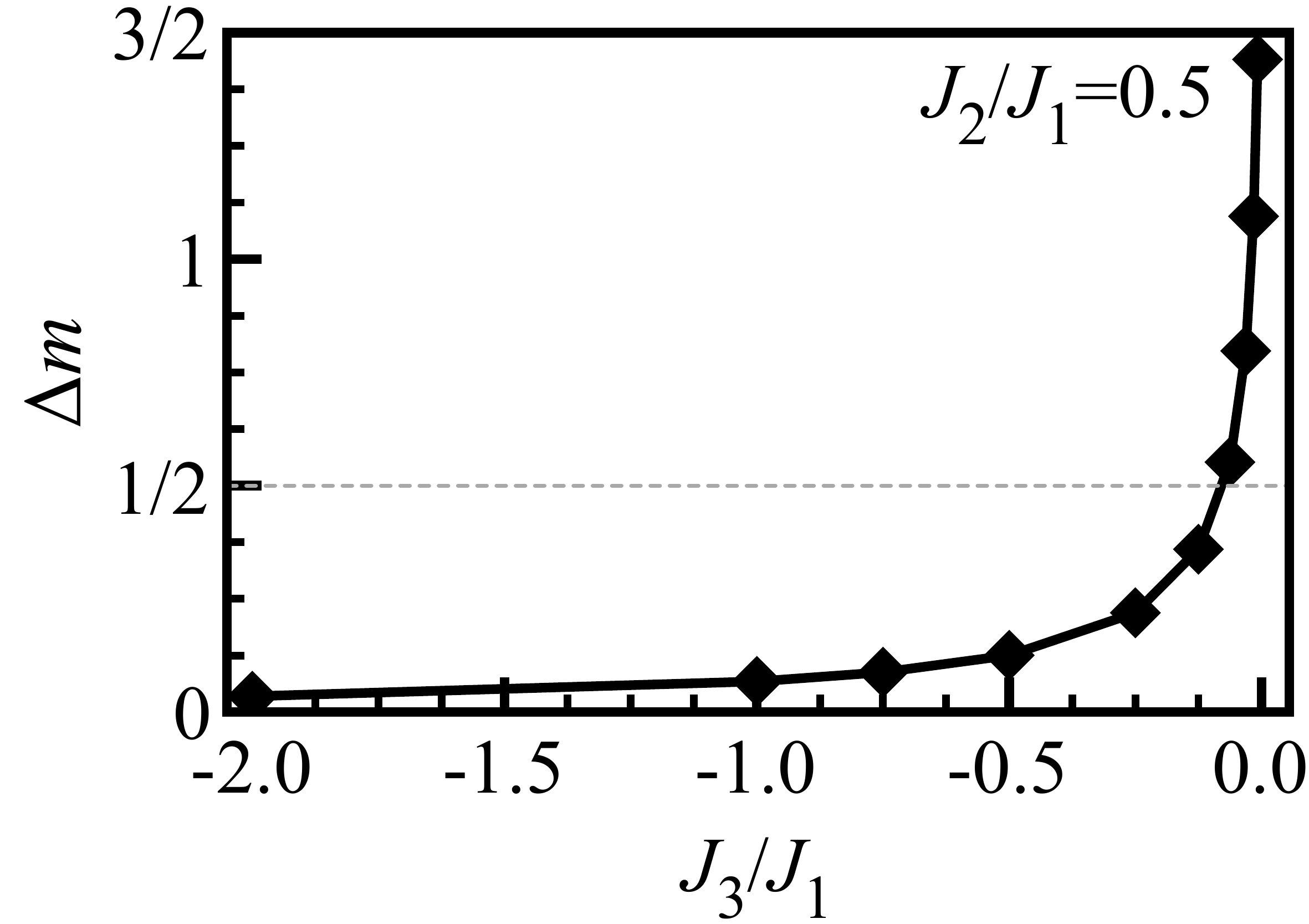}
\caption{Left: Zero point energy due to spin wave excitations as a function of $J_3$. Right: Correction to the N\'eel moment as a function of $J_3$.}
\label{fig:energycrctn}
\end{figure}

\begin{figure}[htbp]
\centering
\includegraphics[width=8.5cm]{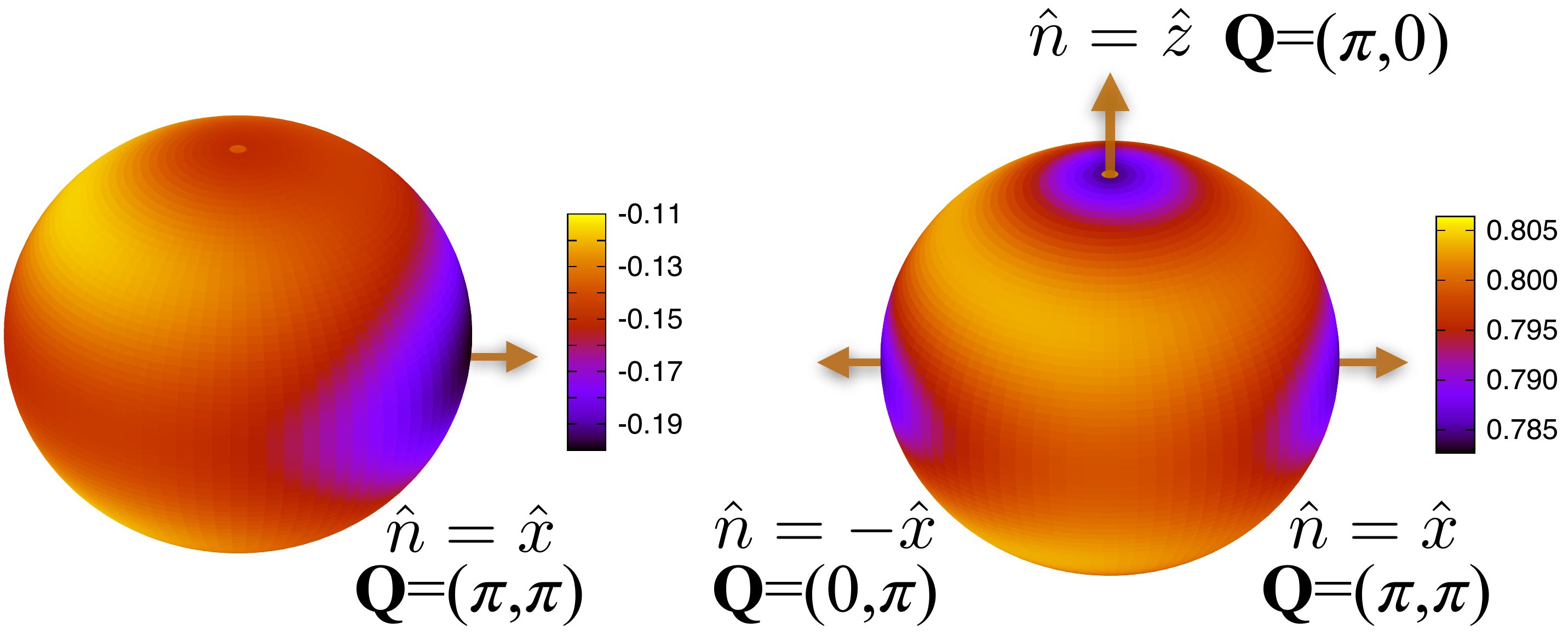}
\caption{Order by disorder due to spin wave fluctuations. Left: Zero point energy. Right: free energy due to spin waves as a function of $\{\theta$,$\varphi\}$. The energy and free energy are plotted on the surface of the Bloch sphere, with the polar angle given by $\theta$ and azimuthal angle given by $\varphi$. The correction to energy is minimum for the N\'eel state, corresponding to $\{\theta,\varphi\}=\{\pi/2,0\}$ or the $\hat{x}$ direction. The free energy is minimum at three points: $\{\theta,\varphi\}=\{0,0\}$ corresponding to horizontal stripe, $\{\theta,\varphi\}=\{\pi/2,\pi\}$ corresponding to vertical stripe and $\{\theta,\varphi\}=\{\pi/2,0\}$ corresponding to N\'eel orders.}
\label{fig:energyfreeenergy}
\end{figure}

At zero temperature, the spin wave Hamiltonian gives an $\mathcal{O}(S)$ correction to the ground state energy: $\Delta E = \sum_{\mathbf{k}} {}^{'} \sum_{j=1}^4 \left\{ \epsilon_{j,\mbk} +  c_\mbk\right\}$. This can be interpreted as zero point energy due to spin wave fluctuations.
In Fig.~\ref{fig:energycrctn}(left), the zero point energy is plotted as a function of $J_3$ for the four classical ground states shown in Fig.~\ref{fig:spinwavedisp}. The N\'eel state has the lowest energy as shown. Indeed, the N\'eel state has the lowest zero point energy among all ground states for any $J_3 <0$. This is illustrated in Fig.~\ref{fig:energyfreeenergy}(left) 
which plots $\Delta E$ for a particular value of $J_3$ ($J_3 = -J_1$) as a function of $\theta$ and $\varphi$ on the surface of the $\hat{n}_{\{\theta,\varphi\}}$ Bloch sphere. 
Thus, with quantum spins at zero temperature, we expect the $(J_2 = J_1/2, J_3<0)$ line to show N\'eel order. We confirm this expectation for the case of $S=1/2$ in Sec.~\ref{sec.ED} using exact diagonalization. 

While the N\'eel state has the lowest energy, it may be destabilized for small $S$ values by quantum fluctuations. The N\'eel ordered-moment has a $1/S$ correction given by $\Delta m = \frac{1}{4N_{\boxtimes}}\sum_{\mbk}\sum_i\langle a_{i,\mbk}^\dagger a_{i,\mbk}\rangle$. When $\Delta m \sim S$, we may surmise that N\'eel order becomes unstable. We plot $\Delta m $ as a function $J_3$ in Fig.~\ref{fig:energycrctn}(right). For the extreme quantum limit of $S=1/2$, we see that the N\'eel state is stable for $J_3 \lesssim -0.1 J_1$.  For weaker $J_3$ couplings, quantum fluctuations destabilize the N\'eel state -- this is consistent with the expectation of a quantum disordered state at $(J_2 = J_1/2, J_3=0)$. 

\subsection{Thermal order by disorder}
\label{sec.thermal}
At finite temperatures, low energy spin wave excitations will contribute to the entropy of the system. In the classical limit, it is the entropy that breaks the degeneracy of the ground state manifold. For classical spins at low temperatures,
the free energy is given by $F=\sum_{\mbk} \sum_i \ln(\epsilon_{i,\mbk})$. The spin wave energies $\epsilon_{i,\mbk}$ here are the same as those obtained by the Holstein Primakov method. Even though the Holstein Primakov method is designed for quantum spin-$S$ spins, it gives the same spectrum as a purely classical derivation using equations of motion. 

We plot the free energy as a function of $\theta$ and $\varphi$ in Fig.~\ref{fig:energyfreeenergy}(right). The effect of thermal fluctuations is very different from that of quantum fluctuations. The lowest free energy occurs in three different states: N\'eel, vertical stripe and horizontal stripe states. Thus, the classical spin model, at zero temperature, breaks global spin rotational symmetry as well as a $Z_3$ symmetry, corresponding to a choice among N\'eel, horizontal stripe and vertical stripe orders. 
At any non-zero temperature, spin rotational symmetry is restored, in line with the Mermin Wagner theorem. However, the discrete $Z_3$ symmetry may survive upto some critical temperature. In section~\ref{sec:classicalMC}, we confirm this picture using Monte Carlo simulations. Our study provides an interesting example where thermal fluctuations and quantum fluctuations give rise to different behaviours. While this is not surprising, there are very few such examples reported in literature~\cite{Bernier2008,Ganeshthesis,Seabra2016}.

\section{Classical Monte Carlo}
\label{sec:classicalMC}
Spin wave theory suggests that the classical spin model should have a finite temperature phase transition above which $Z_3$ symmetry is restored. The $Z_3$ transition in two dimensions is known to be a continuous transition with well established critical exponents. To verify this, we have performed classical Monte Carlo simulations using standard single flip Metropolis and energy conserving microcanonical moves. The simulations were performed on $L \times L$ lattices with periodic boundary conditions, with $L$ upto 120. Focussing on the $J_2 = J_1/2$ line, we simulated many negative $J_3$ values. Starting from random initial configurations, we performed 5$\times$10$^5$ Metropolis moves, with each Metropolis move followed by 3-4 energy conserving microcanonical moves. The first 5$\times$10$^4$ moves were discarded for measurements to allow for equilibration. For each temperature value, we used 10-20 instances to average physical quantities. 
 
\begin{figure}[htbp]
\centering
\includegraphics[width=0.51\textwidth]{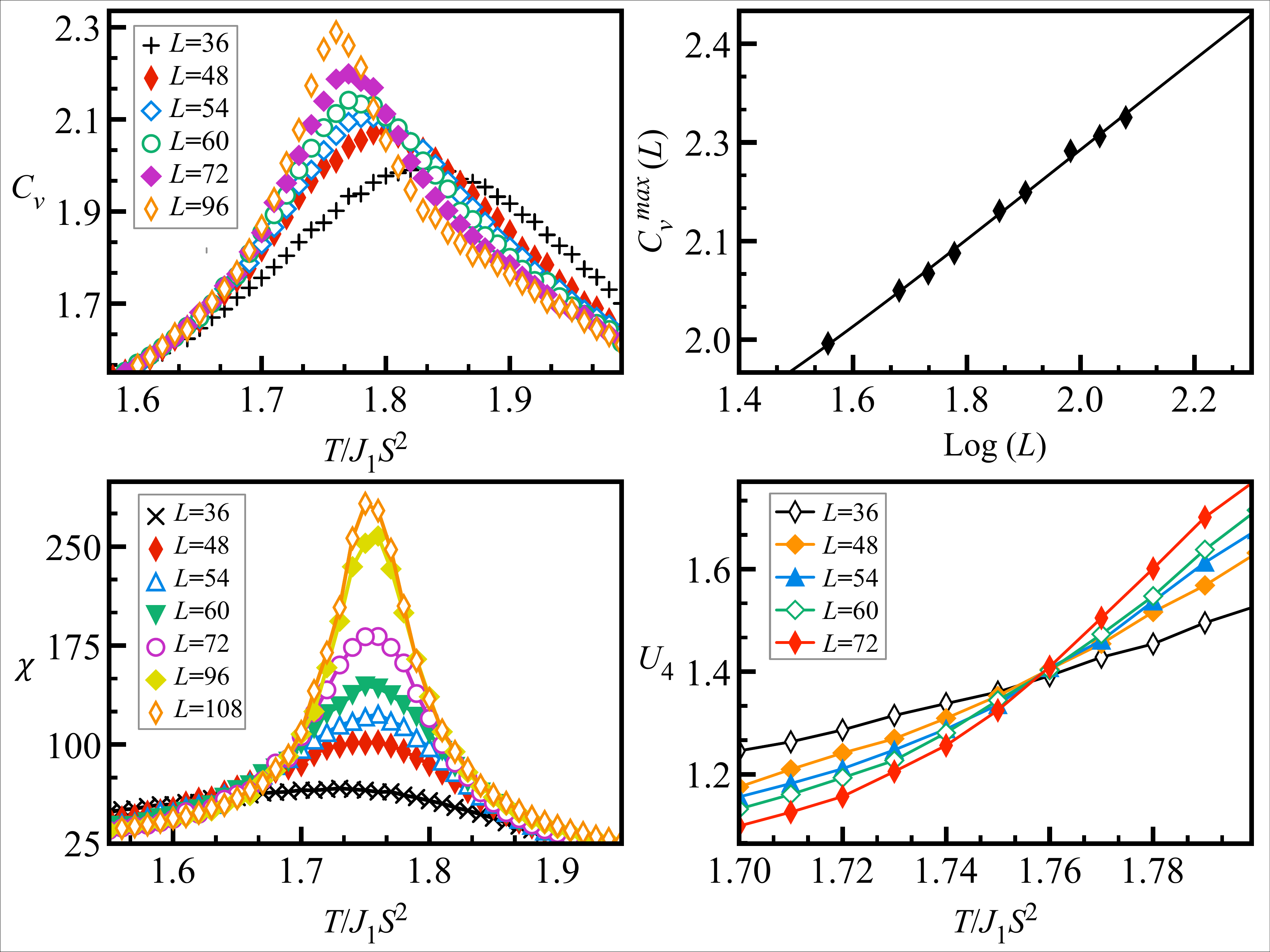}
\caption{Top-left: Specific heat as a function of temperature for different system sizes. Top-right: Specific heat maxima plotted as a function of $\log (L)$ along with the three parameter fit line. Bottom-left: Susceptibility vs. temperature for various system sizes. Bottom-right: The Binder cumulant as a function of temperature. All panels show data for $J_2=J_1/2$ and $J_3=-2J_1$.}
\label{fig:Cv_Cvmax_Xi_BinderCvs.T}
\end{figure}

We compute the specific heat defined by  $C_v=\frac{N}{T^2}(\langle E^2 \rangle-\langle E \rangle^2)$, where $N$=$L^2$. It shows a maximum which grows and shifts with increasing system size, as shown in Fig.~\ref{fig:Cv_Cvmax_Xi_BinderCvs.T}(top-left). This clearly indicates a phase transition, most likely continuous~\cite{Ferdinand1969,Weber2003,Capriotti2004,Mulder2010}. The maximum of specific heat as a function of system size fits well to $C^{max}_v (L)=c_0+c_1 \log (L) +c_2/L$~\cite{Ferdinand1969}. The specific heat maxima along with the fit line are shown in Fig.~\ref{fig:Cv_Cvmax_Xi_BinderCvs.T}(top-right). This further supports a continuous phase transition.

We introduce a local complex order parameter in each square plaquette, following a similar definition on the honeycomb lattice~\cite{Mulder2010},
\begin{eqnarray}
\nonumber
\psi_\n&=&(\hat \Sbf_1.\hat \Sbf_3 +\hat \Sbf_2.\hat \Sbf_4)+\omega(\hat \Sbf_1.\hat \Sbf_2+\hat \Sbf_3.\hat \Sbf_4 )\\
&+&\omega^2(\hat \Sbf_1.\hat \Sbf_4+\hat \Sbf_2.\hat \Sbf_3),
\end{eqnarray}
 where $\omega=e^{i2\pi/3}$, and $(1,2,3,4)$ are labels for spins on a square plaquette with the diagonals being ($1,3$) and ($2,4$), see Fig.~\ref{fig.squares}(left). 
 The order parameter is designed to be proportional to $1$, $\omega$ and $\omega^2$ for N\'eel, horizontal stripe and vertical stripe, respectively.
The average order parameter is defined as $m=\frac{1}{N} \sum_\n \psi_\n$, where $\n$ sums over all square plaquettes in the system.

Signatures of the phase transition are also seen in susceptibility and in the Binder cumulant, defined as $\chi=\frac{N}{T}(\langle  \vert m \vert^2 \rangle-\langle \vert m \vert \rangle^2)$ and $U_4=\langle \vert m \vert^4\rangle/\langle \vert m \vert^2\rangle^2$, respectively. The susceptibility shows a maximum which increases with system size, shown in Fig.~\ref{fig:Cv_Cvmax_Xi_BinderCvs.T}(bottom-left). Fig.~\ref{fig:Cv_Cvmax_Xi_BinderCvs.T}(bottom-right) shows the Binder cumulant which exhibits a crossing, indicative of a continuous transition.

Near a $Z_3$ thermal transition in two dimensions, the specific heat, susceptibility and the order parameter are known to scale as $C_v\propto L^{\alpha/\nu}$, $\chi \propto L^{\gamma/\nu}$, and $\langle|m|\rangle\propto L^{-\beta/\nu}$ respectively, with $\alpha/\nu$=2/5(=0.4), $\beta/\nu$=2/15($\approx$0.1333) and $\gamma/\nu$=26/15($\approx$1.7333)~\cite{Baxterbook}.
 For $(J_2 = J_1/2, J_3=-2J_1)$, we find $T_c/J_1\approx1.75\pm0.01$. The critical exponents are found to be $\alpha/\nu\approx0.402$, $\beta/\nu\approx0.132$ and $\gamma/\nu\approx1.561$, in good agreement with the $Z_3$ (3-state Potts) universality class. 

We point out an important aspect here -- we only see the $Z_3$ transition for $J_3\lesssim -3 J_1/2$. For weaker $J_3$, we do find a broad maximum in specific heat and susceptibility.  However, we do not see clear finite size scaling expected for a phase transition. This can be rationalized in the following way. Spin wave results tell us that at low temperature, there are three states with minimum free energy. The system will break this threefold symmetry and pick one of the three. As we increase temperature, we may expect a $Z_3$ transition \textit{if}  other competing states from outside the threefold set are not accessible to the system. In our system, the other states that could become accessible are the spiral states that are ground states for $J_3 = 0$ but not for $J_3 <0$, see Fig.~\ref{fig:classminima}. These states lie above the three low energy states ($\mathbf{Q} = (\pi,\pi)$, $(\pi,0)$ and $(0,\pi)$), separated by an energy cost proportional to $J_3$. As long as the temperature is below $\sim J_3$, we expect these states to be inaccessible, thereby making way for a $Z_3$ transition. This condition is satisfied for $J_3\lesssim -3 J_1/2$, where we find $T_c \lesssim |J_3|$. When $J_3\gtrsim -3 J_1/2$, we find a broad maximum at some $T_{max.}\gtrsim |J_3|$. Thus, there is a tendency towards a $Z_3$ transition; however, at this temperature, other states are accessed by the system destroying the $Z_3$ character. This is consistent with our expectation that there should be no $Z_3$ transition at $J_3 = 0$. 

\section{Quantum S=1/2 limit at $J_2 = J_1/2,J_3<0$} 
\label{sec.ED}
The $J_1$-$J_2$ has been extensively studied in the quantum $S=1/2$ limit~\cite{Poilblanc1991,Schulz1996,Oitmaa1996}. We are interested in the regime $(J_2 = J_1/2, J_3 < 0)$. Our calculations establish the phase diagram with high certainty and highlight several interesting features. Hitherto, this regime has only been explored using self-consistent spin-spin Green's functions \cite{Mikheyenkov2012} -- our results show that the reported phase diagram misses several important qualitative features.
\subsection{Exact diagonalization}
\label{subsec:ED}

To study the $S=1/2$ limit, we use Lanczos numerical diagonalization in the $S_z=0$ sector, making use of translational symmetries.  We have performed the calculation on $L$=16, 20, 32 and 36 sites clusters with periodic boundary conditions. The quantity of interest is the magnetic order parameter in the ground state, defined as
\begin{eqnarray}
m^2_{s}(\bf{Q})&=& \frac{1}{L^2}\sum_{i,j} \langle \Sbf_i.\Sbf_j \rangle e^{i\Qbf.(\r_i-\r_j)}.
\label{eq:Ms_ord}
\end{eqnarray}
For the N\'eel phase, we have $\Qbf=(\pi,\pi)$. For the stripe phase, we may have $\Qbf=(\pi,0)$ or $\Qbf=(0,\pi)$.  If the computed order parameter extrapolates to a positive value in the thermodynamic limit, we infer that the ground state is ordered. 

\begin{figure}[htbp]
\centering
\includegraphics[width=7.6cm] {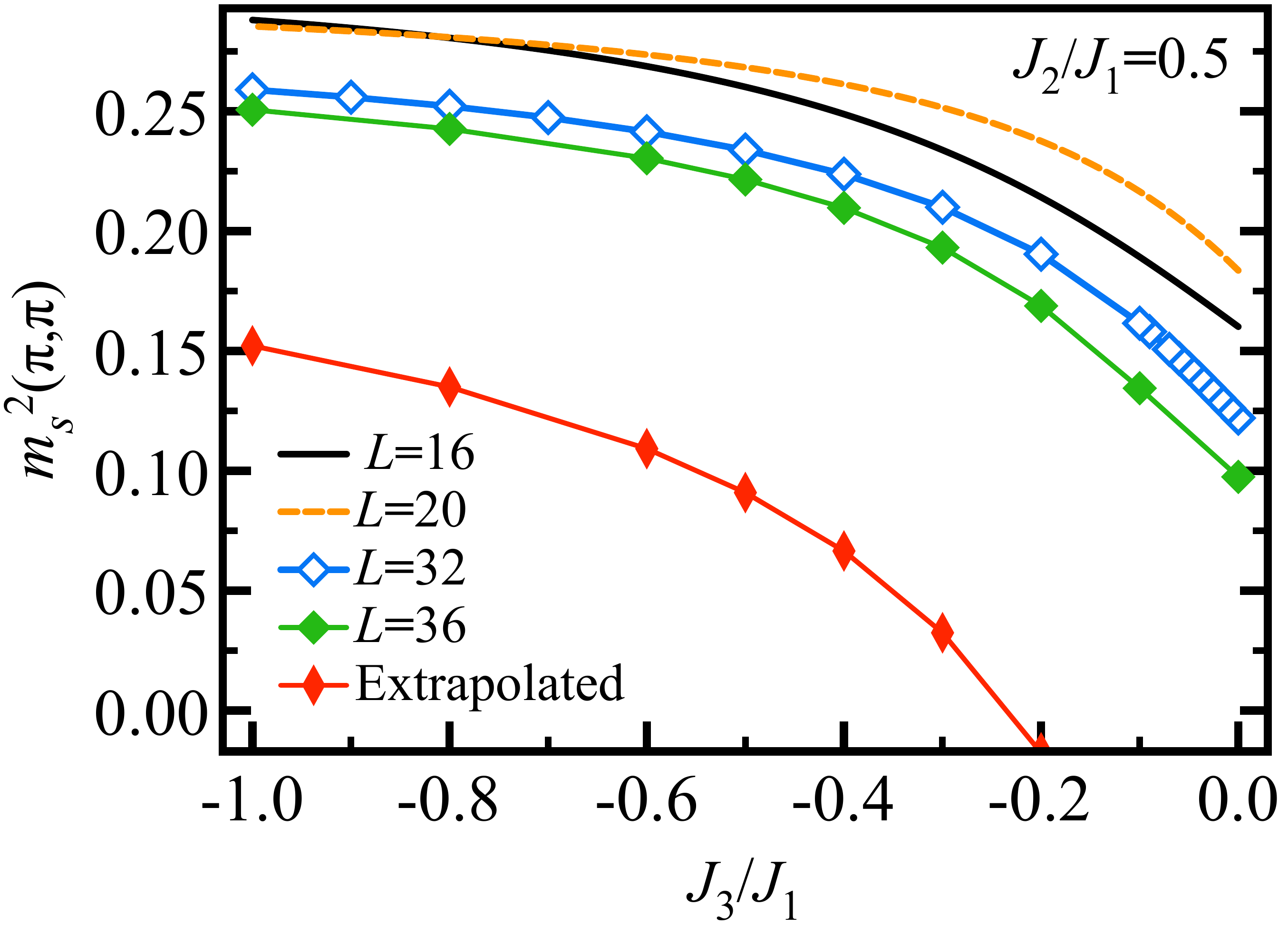}
 \includegraphics[width=7.6cm]{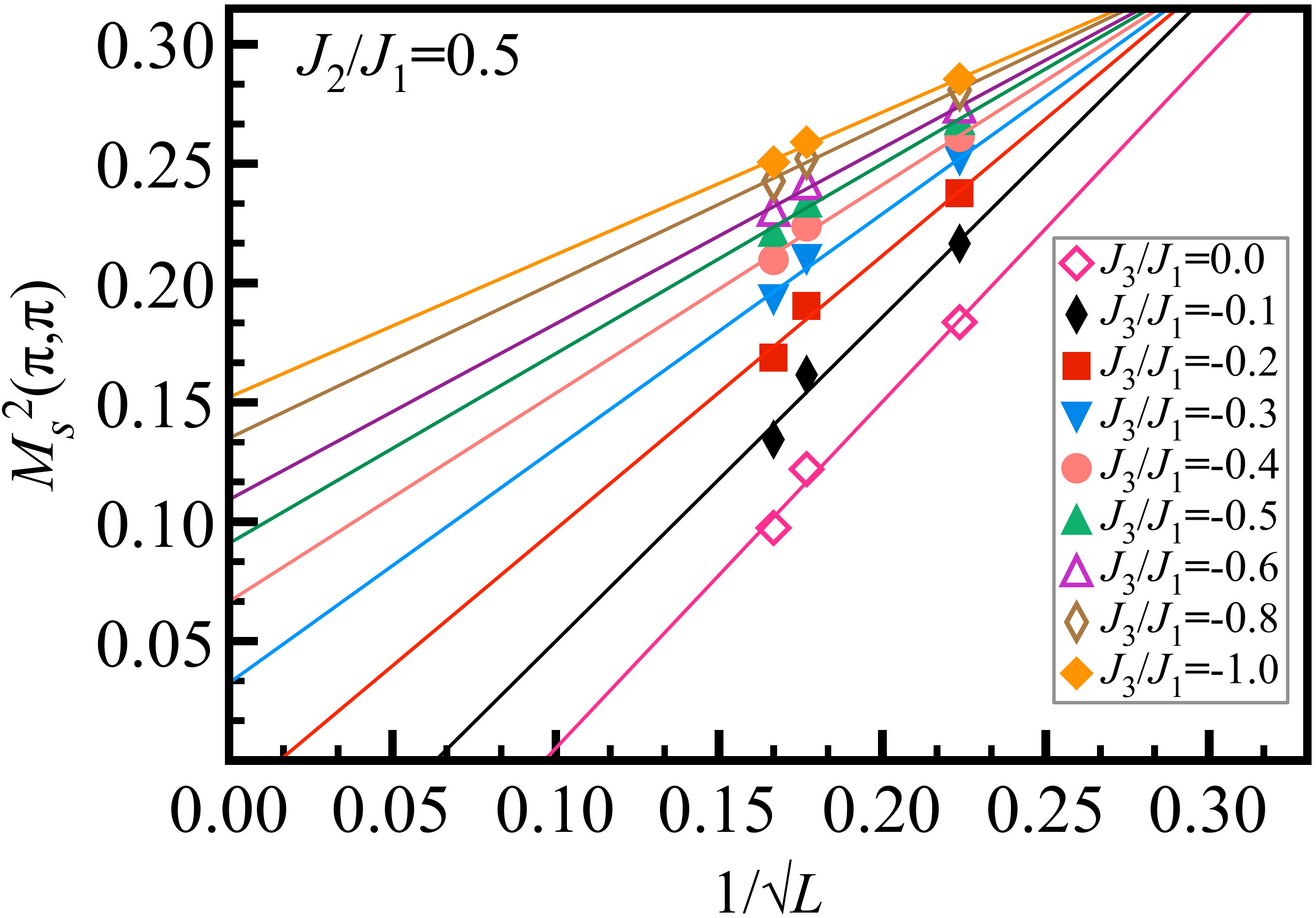}
\caption{ Top: $m^2_{s}(\bf{\pi,\pi})$ plotted as a function of $J_3/J_1$ at $J_2/J_1=0.5$ for $L$=16, 20, 32 and 36. The extrapolated results are from L=20, 32 and 36 clusters. Bottom: Finite size scaling results for $m^2_{s}(\bf{\pi,\pi})$ as function of $1/\sqrt L$. The lines are least-squares fits for the data from $L$= 20, 32 and 36 clusters with the Eq.~\ref{eq:Ms_Nord_Lifnty}.}
 \label{fig:Ms_vs_1bysqrtL}
\end{figure}

Lanczos results for $m^2_{s}(\bf{Q})$ at $\Qbf=(\pi,\pi)$ with ferromagnetic $J_3$ are shown in Fig.~\ref{fig:Ms_vs_1bysqrtL}(top).  We  clearly see that the N\'eel moment increases with increasing (negative) $J_3$.  
To see the phase boundary between the disordered quantum paramagnetic phase and the ordered N\'eel phase, we perform finite size scaling of the Lanczos results. Curiously, the 16 sites cluster does not allow for good finite size scaling, as can be seen in Fig.~\ref{fig:Ms_vs_1bysqrtL}(top). This has also pointed out by Schulz {\it et al} for $J_2/J_1$ around 0.5 and $J_3/J_1=0$~\cite{Schulz1996}; a possible reason is that the 16-site cluster at $J_2=0$ corresponds to a hypercube in four dimensions. We have performed finite size scaling with data from $L$=20, 32 and 36 sites. The data for $m^2_{s}$($\pi,\pi$) scale as~\cite{Schulz1996,Sandvik1997}
\begin{eqnarray}
M^2_{s}(\bf{Q})&=&m^2_{s}(\bf{Q})+\frac{\it{const}}{\sqrt L}.
\label{eq:Ms_Nord_Lifnty}
\end{eqnarray}
The N\'eel moment extrapolated to the thermodynamic limit is shown in Fig.~\ref{fig:Ms_vs_1bysqrtL}(bottom). Our results suggest a non-magnetic quantum paramagnetic ground state for $J_3/J_1$ $\ge-0.2$ along the $J_2 = J_1/2$ line. We see clear evidence for N\'eel order for $J_3/J_1$ $<-0.2$. 

\subsection{Stabilization of N\'eel order in the classical stripe domain}
\label{subsec:stabNeel}
\begin{figure}[htbp]
\centering
\includegraphics[width=0.5\textwidth]{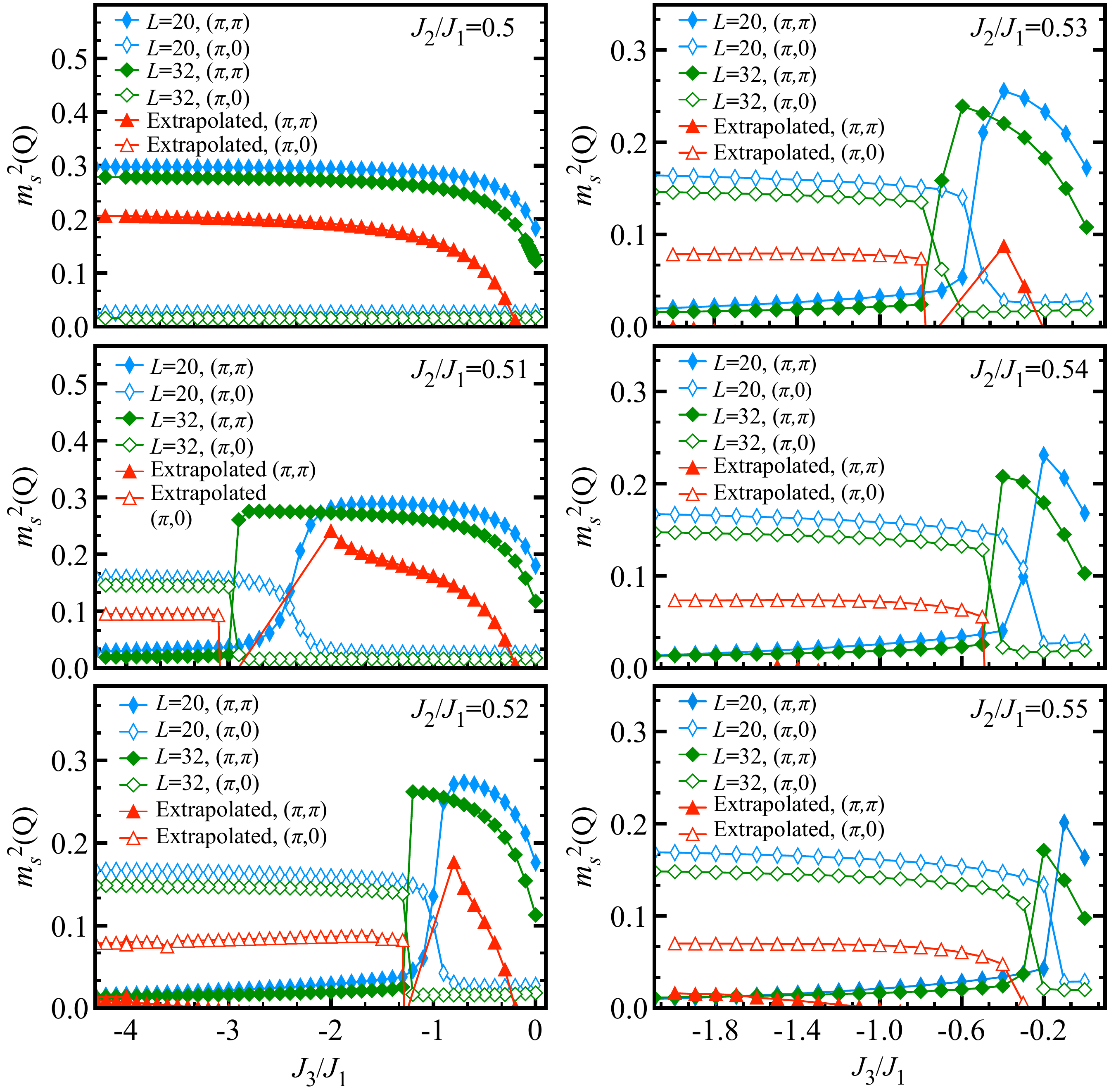}
\caption{$m^2_{s}(\bf{Q})$ for $\Qbf=(\pi,\pi)$ and $(\pi,0)$ along with extrapolated results as a function of $J_3/J_1$ in the range $0.5\le J_2/J_1\le 0.55$. N\'eel order survives within the classical stripe region in a small window around $0.5<J_2/J_1\lesssim 0.53$.}
\label{fig:Ms_vs_J3J251_52_53_54_55}
\end{figure}

Along the $(J_2 = J_1/2, J_3<0/J_1)$ line, the classical ground state is highly degenerate encompassing N\'eel and stripe orders. However, as we have shown at large S (Holstein Primakov spin wave theory) and at $S=1/2$ (exact diagonalization), quantum fluctuations select N\'eel order. This indicates that the N\'eel state has maximal energy lowering from quantum fluctuations. If we increase $J_2$ away from this line, we enter the stripe domain in which the stripe phase has a lower ground state energy than the N\'eel state. However, when we take into account quantum fluctuations, N\'eel order may win over the stripe state as it has greater energy gain from quantum fluctuations. By this reasoning, we expect that the N\'eel state will be stabilized inside the stripe domain -- atleast within a small window close to the $(J_2 = J_1/2, J_3<0)$ line. Indeed, exact diagonalization results confirm this picture. Fig.~\ref{fig:Ms_vs_J3J251_52_53_54_55} shows the obtained values of N\'eel and stripe moments as a function of $J_3$ for different values of $J_2$. We have plotted the magnetic moments for different system sizes along with the values extrapolated to the thermodynamic limit. Interestingly, we find that up to $J_2/J_1$$\approx$0.53, the line $J_3/J_1$ $\gtrsim-0.2$ is a phase boundary between a disordered quantum paramagnetic phase and the ordered N\'eel phase.  We also observe that for $0.5<J_2/J_1\le0.53$, the N\'eel  phase vanishes for large negative $J_3$ depending upon the $J_2/J_1$ ratios.  
For instance, at $J_2/J_1 = 0.51$, we conclude that a paramagnetic phase exists for $0> J_3/J_1 > -0.2$, N\'eel order exists for $-0.2> J_3/J_1 > -2$ and stripe order occurs for $J_3/J_1 < -3.1$. However,  for $0.5<J_2/J_1\le0.53$ with large negative $J_3$, we cannot discern the nature of the transition from N\'eel to stripe order from our finite size numerics.
For example, for $ L=$ 20 and 32 in~Fig.~\ref{fig:Ms_vs_J3J251_52_53_54_55}, there is no consistent pattern in the data points around the N\'eel to stripe transition. 
The 32 site cluster alone seems to indicate a direct first order transition from N\'eel to stripe order; this may indeed hold true in the thermodynamic limit. It is also conceivable that a spin liquid phase may occur within a small window, intervening between the magnetically ordered phases. For $J_2/J_1\gtrsim0.54$ and $J_3/J_1<-0.2$, we find a clear first order transition from the quantum paramagnetic phase to the stripe phase. 
\begin{figure}[htbp]
\centering
\includegraphics[width=7.6cm]{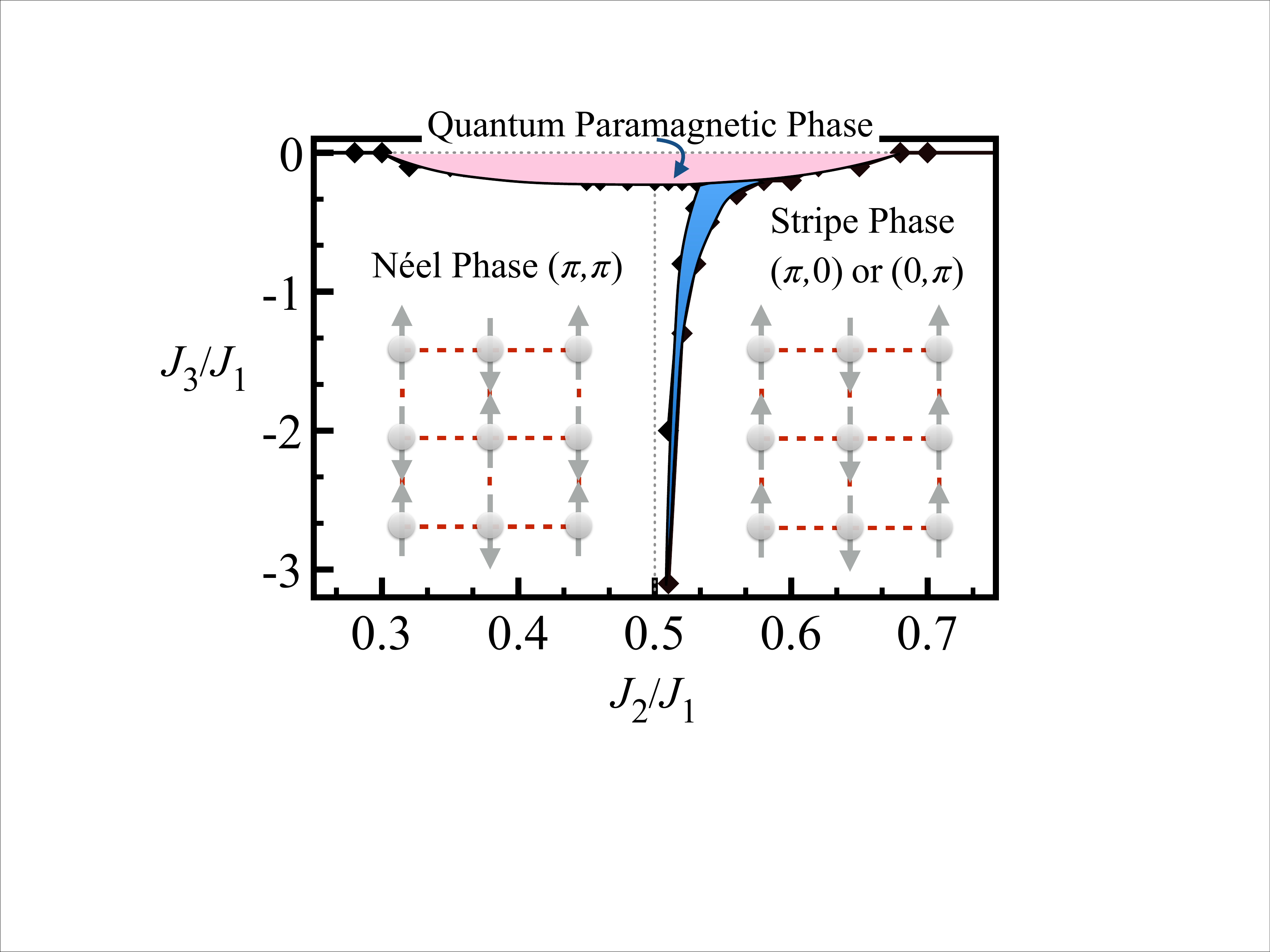}
\caption{Phase diagram in the $S=1/2$ limit, obtained by exact diagonalization. The dashed line at $J_1 = J_1/2$ is the classical phase boundary between N\'eel and stripe order. We cannot determine the nature of the ground state within the blue region, based on our finite size data. }
\label{fig:quantumphaseJ2vsJ3}
\end{figure}

Performing the same analysis at different $J_2$ values, we map out a quantum phase diagram in $J_2$-$J_3$ space as shown in Fig.~\ref{fig:quantumphaseJ2vsJ3}. For, $0.3< J_2/J_1<0.68$ and $0\ge J_3/J_1\ge-0.2$ the ground state is a non magnetic quantum paramagnet (see the pink shaded region in Fig.~\ref{fig:quantumphaseJ2vsJ3}) consistent with the $J_1$-$J_2$ model. We cannot conclusively determine the nature of the ground state within the blue shaded region shown in Fig.~\ref{fig:quantumphaseJ2vsJ3}. The most exciting aspect of this phase diagram is the stabilization of N\'eel order within a small window in the classical stripe domain -- between the dashed line and the blue shaded region in the figure. 

\subsection{Variational plaquette ansatz}
\label{subsec:variational}
The classical model and the quantum model at large-$S$ both possess a four site magnetic unit cell. This suggests that the quantum disordered state at small $S$ and weak $J_3$ coupling may also have a four-site unit cell. With this motivation, we study the $S=1/2$ limit with a plaquette-factorized variational wavefunction:
\begin{equation}
\vert \Psi_{var} \rangle \equiv \prod_{plaq.} \vert \Psi_{plaq.} \rangle.
\end{equation}
The product is over alternate squares -- the shaded squares in Fig.~\ref{fig.squares}(right). As the Hilbert space of a single plaquette is $2^4 = 16$ dimensional, we have 31 real variational parameters after accounting for normalization. We determine  $\vert \Psi_{plaq.} \rangle$ by minimizing the expectation value of the Hamiltonian $\langle \Psi_{var} \vert H_{J_1,J_2,J_3} 
\vert \Psi_{var} \rangle$ by simulated annealing. We denote the minimum energy state by $\vert \Psi_{plaq.} \rangle \equiv \vert 0 \rangle$. 

\begin{figure}[htbp]
\centering
\includegraphics[width=0.5\textwidth]{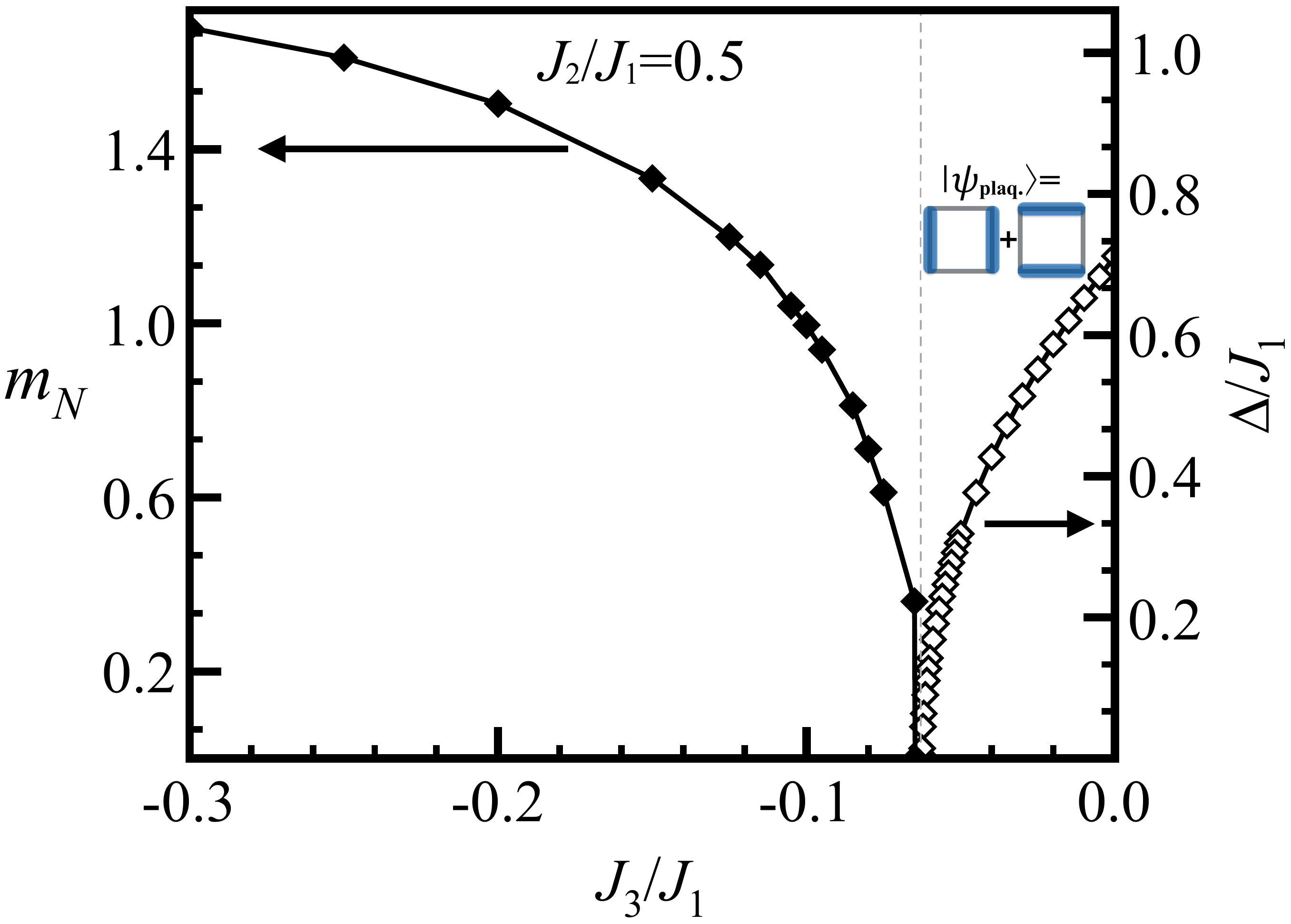}
\caption{Variational wavefunction results: for $J_3 \gtrsim -0.065 J_1$, we have an $s$-wave plaquette-RVB state. Its spin gap is plotted with open diamonds. For $J_3 \lesssim -0.065 J_1$, we have a non-zero N\'eel moment plotted with closed diamonds. The N\'eel moment arises exactly where the spin gap closes in the singlet phase. }
\label{fig:neelplaquatteJ2vsJ3}
\end{figure}

For $0 > J_3 \gtrsim -0.065$, the variational ground state is a singlet with $s$-wave symmetry. When the strength of the $J_3$ coupling is increased beyond $J_3\sim-0.065$,  N\'eel order starts to develop as shown in Fig.~\ref{fig:neelplaquatteJ2vsJ3}. The N\'eel moment is defined as $m_N = \vert  \mathbf{S}_1 - \mathbf{S}_2 + \mathbf{S}_3 - \mathbf{S}_4 \vert$. 
The smooth increase of the N\'eel moment is due $\vert \Psi_{plaq.} \rangle$ acquiring  a triplet component, thus falling within the paradigm of triplon condensation. To further support the triplon condensation picture,  we use a plaquette operator approach to find the spin gap in the $s$-wave singlet phase. 

Having found $\vert 0 \rangle$, the plaquette wavefunction that minimizes the variational energy, we construct the remaining 15 states of the plaquette Hilbert space. We carry out a plaquette-operator analysis taking these 15 states to be excitations that live on plaquette sites. We introduce a bosonic representation with $\vert \ell \rangle_i \equiv b_{i,\ell}^\dagger \vert - \rangle$, where $\vert - \rangle$ represents an unphysical vacuum state with no bosons. The bosonic operator $ b_{i,\ell}^\dagger$ creates the state indexed by $\ell=0,\ldots,15$ at plaquette $i$.
The plaquette-factorized state is captured by taking the $\ell=0$ boson to be condensed. To determine the condensate amplitude, we first consider the single occupancy constraint required of a true representation of the plaquette Hilbert space:
\begin{equation}
\sum_{\ell=0}^{15} b_{i,\ell}^\dagger b_{i,\ell} = 1.
\end{equation}
To satisfy this constraint on average, we choose the condensate amplitude to be $ b_{i,0} \sim b_{i,0}^\dagger \sim \sqrt{1 - \sum_{\ell=1}^{15} b_{i,\ell}^\dagger b_{i,\ell}} $.
 
Rewriting the Hamiltonian using these bosonic operators, we have no linear terms as the ground state minimizes the Hamiltonian. We keep only quadratic terms in the bosons, assuming that the bosons are dilute and interactions can be neglected. This certainly holds true in the $s$-wave singlet phase which has a spin gap. Diagonalizing this quadratic Hamiltonian in each momentum sector, we find the quasiparticle energies. We find that lowest quasiparticle energy (the spin gap) occurs at $\mathbf{k}=0$ consistent with a low-lying N\'eel state. This spin gap is plotted as a function of $J_3$ in Fig.~\ref{fig:neelplaquatteJ2vsJ3}. The spin gap closes at $J_3 \sim -0.065 J_1$ heralding triplon condensation. 

\section{Discussion}\label{discussion}
Motivated by the elusive quantum disordered phase in the square lattice $J_1$-$J_2$ model, we have explored the origin of this phase by adding a tuning knob in the form of a $J_3$ coupling. In the classical model, $(J_2 = J_1/2, J_3=0)$ is a special point at which the Hamiltonian can be written as a sum of squares. This leads to a local constraint wherein the spins on each square should sum to zero, giving rise to an infinite degeneracy. Introducing a ferromagnetic $J_3$ forces every alternate square to have the same spin configuration. This brings down the degeneracy to the number of configurations on a single square with zero total spin. 

Equivalently, the ground state degeneracy can be understood from the point of view of spiral states. At $J_2 = J_1/2, J_3=0$, the usual spiral ansatz tells us that all wavevectors on the edges of the Brillouin zone minimize the energy. The resulting classical ground state manifold is composed of two sectors: (i) single spiral states with wavevector anywhere on the edge of the Brillouin zone, and (ii) coexisting spirals formed from $\mathbf{Q}=(\pi,\pi)$, $(\pi,0)$ and $(0,\pi)$. These three spiral wavevectors have the special property that they can coexist to form a legitimate spin state with uniform spin length. Upon adding a ferromagnetic $J_3$, only the Brillouin zone corners survive as minimum energy wavevectors. Interestingly, this restricts the ground state manifold to sector (ii). The resulting ground state manifold is equivalent to a four site magnetic unit cell with repeating squares. With the $J_3$ coupling, we find that both classical and quantum fluctuations lead to ordered states. We thus surmise that the quantum disordered phase in the $S=1/2$ limit is driven by the classical degeneracy of sector (i) alone. This indicates that the square $J_1$-$J_2$ XY model -- which cannot support non-coplanar coexistence states of sector (ii) -- must also have the same paramagnetic phase as the Heisenberg model. Similar equivalence between the Heisenberg and XY ground states has been recently argued for the Kagome lattice~\cite{Lauchi2015}.

With the $J_3$ coupling, we have shown that classical fluctuations lead to a threefold degeneracy with N\'eel and two stripe states. Classical Monte Carlo simulations reveal a clear thermal transition above which $Z_3$ symmetry is restored. Our results suggest an extremely interesting finite temperature phase diagram with two crossovers. In the stripe phase $(J_2 > J_1/2, J_3<0)$, it is well known that a $Z_2$ transition occurs due to two-fold symmetric stripe order. As we approach the $J_2 = J_1/2$ line, N\'eel order becomes degenerate with the stripes, giving rise to a $Z_3$ transition. 
If we move into the N\'eel domain, $(J_2 < J_1/2, J_3<0)$, we expect no thermal transition as spin rotational symmetry is restored at any infinitesimal temperature. Thus, as $J_2$ is decreased from large values, we expect crossovers from $Z_2$ to $Z_3$ transitions from $Z_3$ to no transition. This is an interesting direction for future research. 

Quantum fluctuations also play an interesting role in this problem. Along $J_2 = J_1/2, J_3 <0$ line, they select N\'eel order as we have shown using spin wave theory and exact diagonalization. Quantum fluctuations favour the N\'eel state so much that they stabilize N\'eel order inside the classical stripe region. The quantum phase diagram may also host a spin liquid phase that intervenes between N\'eel and stripe orders. 
Pursuing a four-site variational ansatz for the quantum $S=1/2$ problem, we find a $s$-wave singlet phase stabilized for small $J_3$ values. The same state has been proposed for the $J_1$-$J_2$ problem\cite{Capriotti2000}. It is suggestive that we find this state when we add a $J_3$ coupling.  

We have studied the fine-tuned parameter line of $J_2 = J_1/2$ in the square lattice antiferromagnet. However, our analysis may be of some relevance to materials such as the iron based superconductors, e.g., BaFe$_2$As$_2$, BaFe$_{1.9}$Ni$_{0.1}$As$_2$, etc. Similar spin models have been proposed for pnictides~\cite{Fernandes2014,Liu2012} as well as iron chalcogenides, e.g., FeSe~\cite{J.KGlasbrenner2015}, both of which are well known to have stripe order. A suitable perturbation, such as pressure, may push these materials towards the $J_2=J_1/2$ limit, thereby bringing the N\'eel state into close competition with stripe order. 
\begin{acknowledgments}
We thank Ioannis Rousochatzakis, Yuan Wan and R. Shankar (Chennai) for useful discussions. 
The simulations were carried out on the HPC Nandadevi cluster at The Institute of Mathematical Sciences.
\end{acknowledgments}

\bibliographystyle{apsrev4-1} 
\bibliography{square_J1J2J3}

\end{document}